\documentclass[twocolumn]{aastex62}
\usepackage{graphics,epsf}
\usepackage{amsmath}                
\usepackage{amsfonts}               
\usepackage{amssymb}                
\usepackage{epsfig}                 
\usepackage{appendix}
\usepackage{graphicx}
\usepackage{float}
\usepackage{color, colortbl}
\usepackage{multirow}
\usepackage{colortbl}
\usepackage[para,online,flushleft]{threeparttable}

\newcommand{\erg}{{~\rm erg}}
\newcommand{\yr}{{~\rm yr}}

\newcommand{\AU}{{~\rm AU}}

\begin{document}

\title{The future influence of six exoplanets on the envelope properties of their parent stars on the giant branches}


\author{Ivan Rapoport}
\affiliation{Department of Physics, Technion – Israel Institute of Technology, Haifa 3200003, Israel; ealealbh@gmail.com; soker@physics.technion.ac.il}

\author{Ealeal Bear}
\affiliation{Department of Physics, Technion – Israel Institute of Technology, Haifa 3200003, Israel; ealealbh@gmail.com; soker@physics.technion.ac.il}

\author{Noam Soker}
\affiliation{Department of Physics, Technion – Israel Institute of Technology, Haifa 3200003, Israel; ealealbh@gmail.com; soker@physics.technion.ac.il}
\affiliation{Guangdong Technion Israel Institute of Technology, Guangdong Province, Shantou 515069, China}

\begin{abstract}
We study the evolution of six exoplanetary systems with the stellar evolutionary code \textsc{mesa} and conclude that they will likely spin-up the envelope of their parent stars on the red giant branch (RGB) or later on the asymptotic giant branch (AGB) to the degree that the mass loss process might become non-spherical. 
We choose six observed exoplanetary systems where the semi-major axis is $a_i \simeq 1-2 \AU$, and use the binary mode of \textsc{mesa} to follow the evolution of the systems. In four systems the star engulfs the planet on the RGB, and in two systems on the AGB, and the systems enter a common envelope evolution (CEE). In two systems where the exoplanet masses are $M_p \simeq 10 M_{\rm J}$, where $M_{\rm J}$ is Jupiter mass, the planet spins-up the envelope to about $10\%$ of the break-up velocity. Such envelopes are likely to have significant non-spherical mass loss geometry. In the other four systems where $M_p \simeq M_{\rm J}$ the planet spins-up the envelope to values of $1-2 \%$ of break-up velocity. Magnetic activity in the envelope that influences dust formation might lead to a small departure from spherical mass loss even in these cases. In the two cases of CEE on the AGB the planet deposits energy to the envelope that amounts to $>10\%$ of the envelope binding energy. We expect this to cause a non-spherical mass loss that will shape an elliptical planetary nebula in each case. 
\end{abstract}

\keywords{stars: AGB and post-AGB; planetary systems; stars: winds, outflows; (ISM:) planetary nebulae: general } 

\section{Introduction} 
\label{sec:intro}

There are two types of motivations to study the role of sub-stellar objects (planets and brown dwarfs) in shaping planetary nebulae (PNe), i.e., shaping the outflow from asymptotic giant branch (AGB) progenitors of PNe. We include in this study also rare cases of PNe that red giant branch (RGB) stars might form (e.g., \citealt{Jonesetal2020RGB}). The first motivation is the growing number of observed potential progenitors of PNe (while they are still on the main sequence) with massive planets around them at the appropriate orbits (we list some systems in section \ref{sec:Systems}). 

The second type of motivations is the great success of stellar binary interaction scenarios to account for the shaping of many PNe with the realization that there are not enough stellar binary systems to account for all non-spherical PNe (e.g., \citealt{DeMarcoSoker2011}). The number of studies that support the stellar binary shaping of PNe amounts to several hundreds, and so we limit the list to a fraction of papers from the last five years 
(e.g.  \citealt{Akrasetal2016, Chiotellisetal2016, Jonesetal2016, GarciaRojasetal2016, Hillwigetal2016a, Bondetal2016, Madappattetal2016, Alietal2016, JonesBoffin2017b, Barker2018, BondCiardullo2018, Bujarrabaletal2018, Chenetal2018, Danehkaretal2018, Franketal2018, GarciaSeguraetal2018, Hillwig2018,  MacLeodetal2018b, Miszalskietal2018PASA, Sahai2018ASPC, Wessonetal2018, Brownetal2019, Desmursetal2019, Jonesetal2019triple, Kimetal2019, Kovarietal2019, Miszalskietal2019MNRAS487, Oroszetal2019, Akrasetal2020, Alleretal2020, BermudezBustamanteetal2020, Jacobyetal2020, Jones2020CEE, Jones2020Galax, Mundayetal2020}). 
   
The hard time of planets to survive the evolution of PN progenitors, and the severe difficulties of detecting planets around central stars of PNe in case they do survive, explain the relatively small number of papers that study PN shaping by planets and brown dwarfs (e.g, some papers from the last five years, \citealt{Kervellaetal2016, Boyle2018PhDT, SabachSoker2018a, SabachSoker2018b, Salasetal2019, Schaffenrothetal2019, Decinetal2020}). 
In some cases energy deposition in the RGB or AGB envelope that inflates the envelope to much larger dimensions than what regular evolution does, might help a planet that is engulfed at this phase to survive  (e.g.,  \citealt{Bearetal2011, Bearetal2021, Lagosetal2021, Chamandyetal2021}). 

The challenge for shaping PNe by planets is to find processes by which a companion with a mass of $\approx 1 \%$ of the AGB (or RGB) star influences the mass loss geometry to a detectable level. Stellar companions can affect the envelope of the PN progenitor and its mass loss in some relatively energetic processes. Most notable processes that  three-dimensional hydrodynamical simulations of the common envelope evolution (CEE) with stellar companions have revealed over the years is the inflation of the envelope and ejection of a relatively dense equatorial outflow (e.g., \citealt{LivioSoker1988, RasioLivio1996, SandquistTaam1998, RickerTaam2008,  Passyetal2012, Nandezetal2014, Ohlmannetal2016, Iaconietal2017, MacLeodetal2018a}). A brown dwarf affects the envelope much less than a stellar companion (e.g., \citealt{Krameretal2020}), and a planet will have an even smaller effect of this type. Other stellar-induced energetic processes include, among others, the extreme deformation of the AGB (or RGB) envelope at the end of the CEE to form two opposite polar `funnels' (e.g., \citealt{Soker1992, Reichardtetal2019, GarciaSeguraetal2020, Zouetal2020}), and the launching of jets inside the envelope (e.g., \citealt{Chamandyetal2018a, LopezCamaraetal2019, Schreieretal2019, Shiberetal2019, LopezCamaraetal2020}). 
Planets that experience the CEE with AGB and RGB stars are unable to have these large effects. Even if a planet does launch jets, the result is likely to be only two small opposite clumps along the symmetry axis (\textit{ansae}; FLIERS), e.g., \cite{SokerWorkPlans2020}. 

The influence of planets on the mass loss geometry is via non-linear effects that amplify the perturbations of the planets. In particular, the possible important role of dust formation in determining the mass loss rate and geometry  during the CEE (e.g., \citealt{Soker1992b, Soker1998AGB, GlanzPerets2018, Iaconietal2019, Iaconi2020}). These effects include the action of a dynamo that amplifies magnetic fields in giant stars (e.g., \citealt{LealFerreiraetal2013, Vlemmings2018}) after the planet spins-up the envelope (e.g., \citealt{Soker1998AGB, NordhausBlackman2006}). These magnetic fields might affect the formation of dust on the surface of the giant star, and via that the mass loss geometry (e.g., \citealt{Soker2000, Soker2001a, Khourietal2020}). Another process of a planet deep inside the giant envelope is the excitation of waves whose amplitude becomes large on the giant surface to the degree that they influence the efficiency of dust formation   (e.g., \citealt{Soker1993}). 

We therefore set the goal to examine the degree of spin-up that parent stars of some observed exoplanets will suffer as they swallow their planet, and to calculate how much orbital energy the core-planet system releases relative to the binding energy of the giant envelope. The notion that planets spin-up their parent star on the RGB or AGB is not new of course (e.g., \citealt{Soker1996, SiessLivio1999AGB, Massarotti2008, Carlbergetal2009, Nordhausetal2010, MustillVillaver2012, NordhausSpiegel2013, GarciaSeguraetal2014, Priviteraetal2016I, Priviteraetal2016II, Staffetal2016, AguileraGomezetal2016, Veras2016, Guoetal2017, Raoetal2018, Jimenezetal2020}). Our new contribution is the calculation of these effects in specific observed exoplanetary systems that are likely to shape future PNe. Namely, we only deal with planets that the star swallows on the upper RGB or during the AGB. We do not deal with exoplanets with a semi-major axis smaller than $\simeq 1 \AU$. 

This study follows that by \cite{Hegazietal2020} who determine the potential role of  three exoplanets and two brown dwarfs  on the future RGB or AGB evolution of their parent star. We list two of these systems and four other exoplanets in section \ref{sec:Systems}. Like \cite{Hegazietal2020} we use the Modules for Experiments in Stellar Astrophysics in its binary mode (\textsc{MESA-binary}) as we describe in section \ref{sec:method}. A key ingredient in the study of \cite{Hegazietal2020} that we adopt here is the usage of a lower than traditional mass loss rate on the giant branches. This particularly affects the stellar evolution on the upper AGB. This lower mass loss rate approach is based on the argument that in a case where no companion, stellar or sub-stellar, spins-up the envelope of an AGB star, the mass loss rate of the AGB star is much lower than what traditional formulas give (\citealt{SabachSoker2018a, SabachSoker2018b}; they termed such stars that suffer no spin-up interaction with a companion angular momentum isolated stars or {\it Jsolated stars}). 
These three earlier studies of our group discuss some aspects that we do not deal with here, like the brightest end of the PN luminosity function in old stellar populations. However, they did not study the degree of spin-up and spiraling-in of the planet into the giant envelope which we study here (section \ref{sec:evolution}).
We summarize our results in section \ref{sec:summary}.

\section{NUMERICAL METHOD}
\label{sec:method}
We use \textsc{mesa-binary} (version 10398; \citealt{Paxtonetal2011,Paxtonetal2013,Paxtonetal2015,Paxtonetal2018,Paxtonetal2019}). We follow the inlist of \cite{Hegazietal2020} (which is based on the example of stellar plus point mass) in order to investigate the impact the planet has on the stellar evolution. We take the planet to be a point mass (i.e., we do not follow its evolution or possible mass accretion by the planet). 
We here list parameters that we change from their default values in the \textsc{mesa-binary} mode for all runs.
\begin{itemize}
    \item We set the initial ratio of the angular velocity ($\omega$) and the critical angular velocity ($\omega_{\rm critical}$) of the primary star, while it is on the main sequence,  in all systems to be $\frac{\omega}{\omega_{\rm critical}}=0.001$. We took this value as an initial rotation for slow rotators. This slow initial rotation has no effect on our study of giant stars.
    \item We allow tidal circulation ($do~tidal~circ$) and synchronization ($do~tidal~sync$) because tidal forces and their effects become very important close to the beginning of the CEE.
\item To avoid numerical difficulties we do not limit the minimum of the time step.
    \item We took into account mass-loss by wind, and we used Reimer's and Blocker's formulas for the RGB and AGB evolutionary stages, respectively. The commonly used value of the wind mass-loss rate efficiency parameter is $\eta \simeq 0.5$. We assume a lower mass loss rate as we discussed above, and take 
    $\eta=0.09$ for HIP~114933~b and $\eta=0.12$ for all other systems. 

    \end{itemize}

 Let us elaborate on the last assumption of $\eta<0.5$. As we mention in section \ref{sec:intro}, this is an assumption that we adopt from earlier works (\citealt{SabachSoker2018a, SabachSoker2018b}). These earlier studies based this assumption on the argument that the stellar samples that have been used to derive $\eta \simeq 0.5$ are contaminated by binary systems that increase the mass loss rate. We take the specific values of $\eta=0.09$ or $\eta=0.12$ as these are about the maximum values of $\eta$ that allow the respective stars to engulf their exoplanets. Larger values of $\eta$ will not bring the respective stars to engulf their planets. 
This assumption clearly needs a future verification.

\section{The sample of observed planetary systems}
\label{sec:Systems}

We search for exoplanetary systems where a massive planet might enter the envelope of their parent star along the RGB or the AGB, i.e., the system enters a CEE. We do not follow the CEE, but we do determine the properties of the giant star at the onset of the CEE. We search for such systems in the Extrasolar Planets Encyclopaedia; (exoplanet.eu; \citealt{Schneideretal2011}).
\cite{Hegazietal2020} already studied the evolution of the two exoplanets HIP~75458~b and beta~Pic~c. However, they did not examine the particular properties of the giants that we are interested in. 
For the systems HIP~90988~b, HIP~75092~b, and HIP~114933~b we took the initial parameters from \cite{Jonesetal2021}, while for the initial parameters of the system HD~222076~b we used the studies of \cite{Jiangetal2020} and \cite{Wittenmyeretal2017}.
In systems with uncertain inclination (HIP~114933~b, HIP~90988~b, and HIP~75092~b in this study) we take $\sin i=0.5$.
We list the six systems we study in Table \ref{tab:Table1}.
\begin{table*} 
\centering
\begin{tabular}{|l||l|l|l|l||l|l|l|l|l||l|l|l|l|}
\hline
 & $M_{\ast,i}$ &  $M_p$ &  $a_i$ &  $e_i$ & CEE & $M_{\rm *,CEE}$ & $R_{\rm CEE}$ & $M_{\rm   env,CEE}$ & $e_{\rm CEE}$ & $\frac{\omega_{\rm   CEE}}{\omega_{\rm tidal}}$ & $\frac{\omega_{\rm   CEE}}{\omega_{\rm critical}}$ & $\frac{E_{\rm orbit}}{E_{\rm env}}$ & $E_{\rm env}$\\ \hline
 & $M_\odot$& $M_{\rm J}$ & $R_{\odot}$ &  &  & $M_\odot$ & $R_{\odot}$ & $M_\odot$ &  &  &  &  & $10^{46} \erg$ \\ \hline
HIP 75458 b & 1.4  & 9.40 & 273.00 & 0.71 & RGB & 1.39 & 38.39  & 1.05 & 0.005 & 2.8 & 0.109 & 0.048  & 12 \\ \hline
HIP 90988 b & 1.3  & 0.98 & 270.82 & 0.08 & RGB & 1.28 & 96.49  & 0.87 & 0.003 & 3.3 & 0.015 & 0.015  & 4.8 \\ \hline
HIP 75092 b & 1.28 & 0.89 & 434.21 & 0.42 & RGB & 1.25 & 138.85 & 0.79 & 0.022 & 3.9 & 0.014 & 0.022  & 3.3 \\ \hline
HD 222076 b & 1.38 & 1.57 & 393.34 & 0.08 & RGB & 1.35 & 140.29 & 0.89 & 0.004 & 3.4 & 0.023 & 0.034  & 3.8 \\ \hline
beta Pic c  & 1.73 & 9.37 & 585.00 & 0.25 & AGB & 1.67 & 231.16 & 1.12 & 0.020 & 4.4 & 0.087 & 0.275  & 3.4 \\ \hline
HIP 114933 b& 1.39 & 0.97 & 610.44 & 0.21 & AGB & 1.18 & 368.34 & 0.61 & 0.154 & 54.8 & 0.011 & 0.106 & 0.94 \\ \hline
\end{tabular}
\caption{The six exoplanetary systems that we study here. 
The first column gives the name of the system, while the next four columns list four properties that are derived from observations, the stellar initial mass $M_{\ast,i}$, the planet mass $M_{p}$, the initial semi-major axis $a_i$, and the initial eccentricity ($e_{i}$). The next four column list the properties of the system at the onset of the CEE, the phase of the parent star (RGB or AGB), its mass ($M_{\rm *,CEE}$), its radius ($R_{\rm CEE}$), and its envelope mass ($M_{\rm env,CEE}$), and the eccentricity of the planet-giant binary system ($e_{\rm CEE}$). The next three columns give three relevant ratios for our study, the ratio of the envelope angular velocities at the end of the CEE ($\omega_{\rm CEE}$) to that at the onset of the CEE ($\omega_{\rm tidal}$),  the ratio of $\omega_{\rm CEE}$ to the breakup angular speed of the giant ($\omega_{\rm critical}$), and the ratio of the orbital energy that the planet releases at the end of the CEE when the orbital separation is $a=1R_\odot$ ($E_{\rm orbit}$) to the binding energy of the envelope that resides above $r=1R_\odot$ ($E_{\rm env}$). For reference we also list this envelope binding energy in the last column. }
\label{tab:Table1}
\end{table*}


In the first five columns in Table \ref{tab:Table1} we list the name of the system/planet and four relevant properties of the exoplanetary systems that observations give, the initial stellar mass, the planet mass, the initial semi-major axis, and the eccentricity. 

\section{Evolution to the CEE}
\label{sec:evolution}

We simulate the evolution of the six systems that we list in Table \ref{tab:Table1}, starting with their initial properties from the table and with metalicity of $z=0.02$. We end the simulations when the planet enters the envelope of its parent star (i.e., the system enters a CEE), either when the star is on the RGB or on the AGB, as we indicate in the sixth column of the table. We record the mass of the star, its radius, and its envelope mass at the onset of the CEE, as well as of the eccentricity of the orbit of the planet at that time. We list these quantities at the onset of the CEE in columns 7-10 of Table \ref{tab:Table1}. 
 
Although we do not simulate the CEE, we do calculate relevant properties. Due to tidal interaction in the pre-CEE phase the planet spins-up the giant envelope to an angular velocity of $\omega_{\rm tidal}$. After it enters the envelope the planet spirals-in to very small radius and further spins-up the giant envelope. Under the assumption that the envelope structure does not change (see below), we calculate the final angular velocity of the envelope (after the CEE) $\omega_{\rm CEE}$. In the eleventh column we list the ratio $\omega_{\rm CEE}/\omega_{\rm tidal}$ and in the twelfth column we list the ratio $\omega_{\rm CEE}/\omega_{\rm critical}$, where $\omega_{\rm critical}$ is the critical (breakup) angular velocity of the envelope above which it breaks apart. 

The above assumption that the envelope structure does not change during the CEE holds when the planet does not deposit much angular momentum or much energy. However, in the cases of HIP~75458~b and beta~Pic~c the envelope angular momentum reaches values of $\omega _{\rm CEE} \simeq 0.1 \omega_{\rm critical}$ and in the cases of beta~Pic~c and HIP~114933~b the orbital energies that the planets release is significant. We do expect the envelope to expand and flatten somewhat in response, increasing the value of the envelope moment of inertia and reduces the angular velocity. Nonetheless, the envelope are substantially deformed in these three cases as well.   

When the planet spirals-in it releases orbital gravitational energy until it is destroyed at a very small orbital separation (e.g., \citealt{Krameretal2020} for a recent study). We calculate the orbital energy $E_{\rm orbit}$ that the system releases from the onset of the CEE to an orbital separation of $a_{\rm post-CEE}=1 R_\odot$. Because the initial orbital separation is much larger than $1 R_\odot$, this energy is $E_{\rm orbit}=0.5 G M_{\ast,1}M_p/1R_\odot$, where $M_{\ast,1}$ is the giant mass inward to $r=1R_\odot$. We calculate the binding energy of the envelope that resides above $r=a_{\rm post-CEE}=1 R_\odot$, $E_{\rm env}$.  In the next-to-last column of the Table \ref{tab:Table1} we list the ratio $E_{\rm orbit}/E_{\rm env}$, and in the last column we list $E_{\rm env}$.   

In calculating the binding energy of the envelope that resides at $r > 1 R_\odot$ we took half the gravitational energy of that part of the envelope $E_{\rm env} = 0.5 \vert U_{\rm grav} \vert$,  where $U_{\rm grav} =-\int^{M_\ast}_{M_{\ast,1}} [G M(r)/r] dm $, and $M_\ast$ is the stellar mass. At a radius of $\approx 1 R_\odot$ the binding energy calculated this way and that calculated by adding the gravitational and internal energy  become equal (e.g., \citealt{Lohevetal2019}).
We take a final orbital radius for the planet at $a_{\rm post-CEE}=1 R_\odot$ because at about that radius massive planets start to evaporate. \cite{Krameretal2020}, for example, list the orbital radius at which their hot giant envelope starts to evaporate a planet of mass $M_p=0.01M_\odot$ as $r_{\rm eva} = 1.27 R_\odot$. The planet might spiral-in further while it is evaporated because its tidal destruction radius is much smaller, $r_{\rm tid} = 0.06R_\odot$. Our results are sensitive to the uncertain value of $a_{\rm post-CEE}$ and therefore there are some uncertainties in the values that we calculate based on the value of $a_{\rm post-CEE}$.  
   
\begin{figure}
\vskip -1.50 cm
 \hskip -2.00 cm
 \includegraphics[scale=0.6]{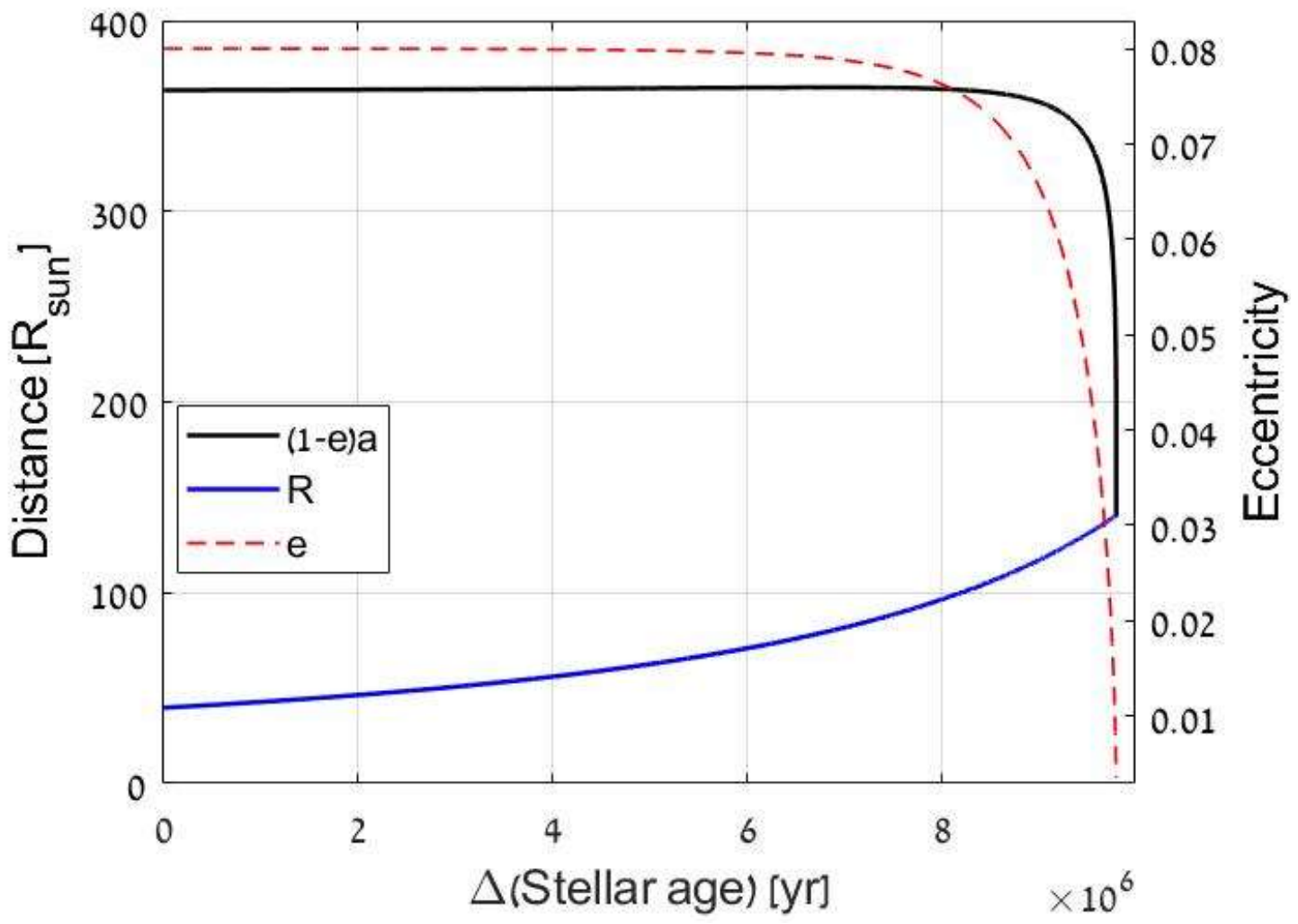}\\
 \vskip -10.00 cm
\caption{ The eccentricity (red-dashed line), stellar radius (blue line) and periastron distance (black line) for HD~222076~b as a function of time starting at $t = 3.74 \times 10^9 \yr$ and ending at the onset of the CEE. 
}
 \label{fig:HD222076graph}
 \end{figure}

\begin{figure}
 \vskip -1.0 cm
 \hskip -2.00 cm
 \includegraphics[scale=0.6]{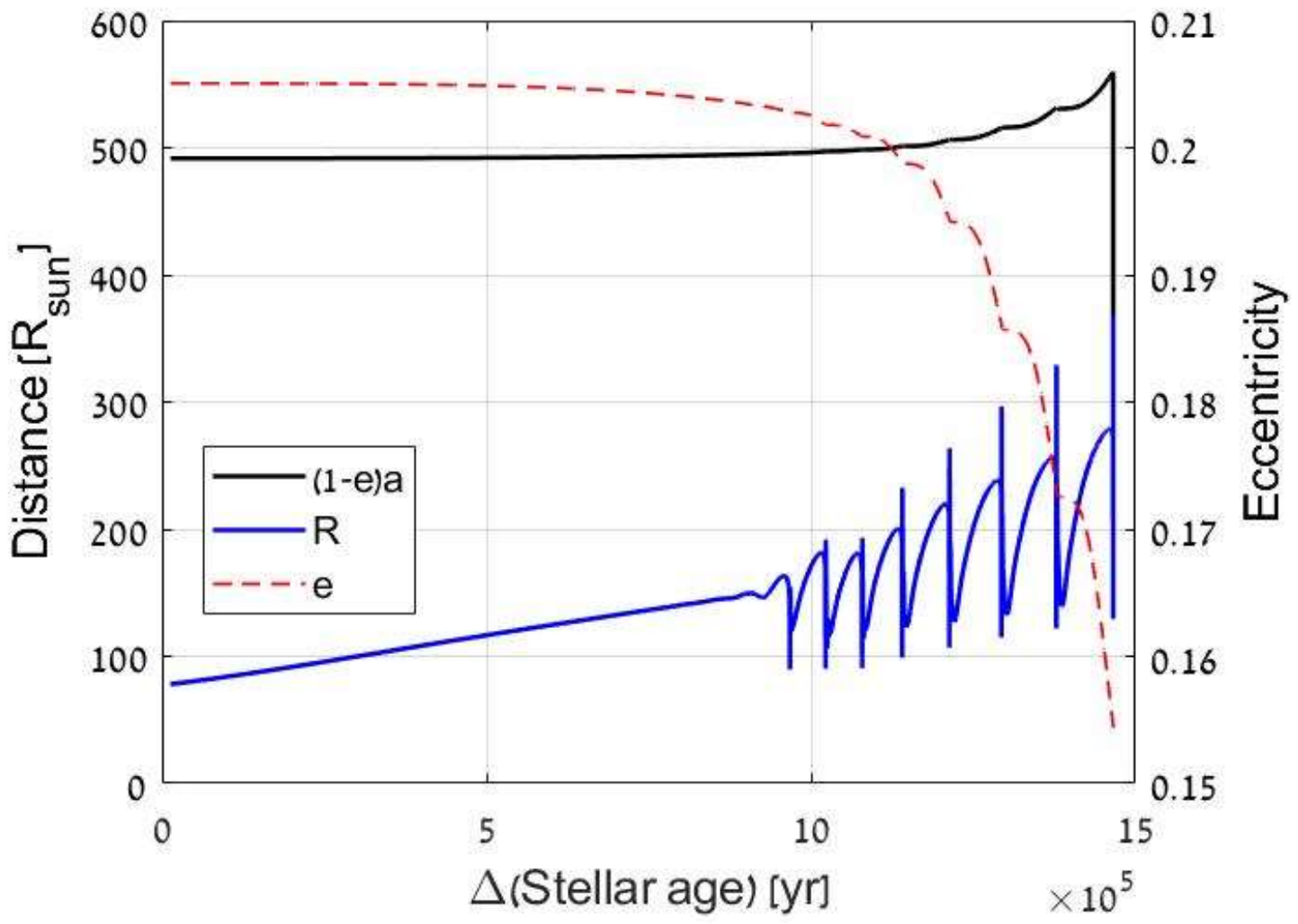}\\
 \vskip -10.00 cm
\caption{Similar to Fig. \ref{fig:HD222076graph}, but for the system HIP~114933~b and starting at $t = 3.77 \times 10^9 \yr$. }
 \label{fig:HIP114933graph}
 \end{figure}

We demonstrate the evolution of two systems in Fig. \ref{fig:HD222076graph} for a planet engulfment on the RGB, and in Fig. \ref{fig:HIP114933graph} for a planet engulfment on the AGB (see \citealt{Hegazietal2020} for the evolution of  HIP~75458~b and beta~Pic~c). 
We present the evolution of the stellar radius (blue solid line), the periastron distance $(1-e)a$, and the eccentricity of these two systems.  

The rapid fall of the planet to the envelope with the increase in stellar radius $R$ demonstrates the well-known sensitivity of the tidal forces to the ratio of $R/a$. Like previous studies (e.g., \citealt{Hegazietal2020}), we find that AGB stars are likely to engulf planets shortly after a helium shell flash (thermal pulse). These helium shell flashes are seen in Fig. \ref{fig:HIP114933graph} as spikes in the stellar radii.

\section{Discussion and Summary}
\label{sec:summary}
 
The degree to which planets can shape the outflow from RGB and AGB stars is an open question. As a result of that, the fraction of PNe that are shaped by planetary systems (and brown dwarfs) is also an open question. The difficulties in answering these questions result from the complicated shaping mechanisms by planets as well as the large difficulties of observations to detect planets in PNe, mainly because many planets do not survive the evolution. 
As such, our view is that many small steps will achieve the goal of answering these questions. The present study is one such small step where we study observed planetary systems. We examine the possible future influence of the planets on the post-CEE giant envelope of their parent star. 
 
Planets might influence the mass loss rate and geometry by inducing other processes (see section \ref{sec:intro}).
One process is the excitation of waves that have large surface amplitudes when the planet is deep in the envelope. Another process is the spinning-up of the giant envelope to a degree that allows the operation of an efficient dynamo. The sun rotates at about half a per cent of its breakup velocity $\omega_{\rm critical}$, and despite this low value has a pronounced dynamo activity to the degree than the wind from the sun is not spherical, neither in velocity nor in mass loss rate. In RGB and AGB stars the magnetic activity might make the geometry of dust formation non-spherical (section \ref{sec:intro}). More dust formation implies a higher mass loss rate from these luminous cool giants.  

We learn from Table  \ref{tab:Table1} that in all cases the planets spin-up the envelopes of their parent stars to angular velocities of $\omega_{CEE} > 0.01 \omega_{\rm critical}$. We expect a non-negligible magnetic activity in all these giants. The spin-up process is particularly significant in the two systems with massive planets, $M_{P} \simeq 10 M_{\rm J}$, where the angular velocity is about ten per cent of the breakup angular velocity.
In these two cases,  HIP~75458~b and beta~Pic~c, the envelope achieves significant rotation before the planet even enters the envelope, and more so after the CEE. 
 
It is not clear whether planets can accrete sufficient amount of mass to launch jets before they enter the envelope. \cite{Salasetal2019} present an optimistic view on planet launching jets in a triple system where a tertiary star keeps the planet outside the envelope and on an eccentric orbit. We also consider another possibility (\citealt{SokerWorkPlans2020} and references therein), according to which the most likely process for planets to form jets is by forming an accretion disk around the core of the RGB or AGB star after the core tidally destroys the planet. All these processes require further studies. 

The direct deposition of significant amount of orbital energy to the envelope takes place only at the end of the CEE. As expected, we learn from the next-to-last last column of Table \ref{tab:Table1} that a large ratio of the orbital energy that the planet releases to the binding energy of the giant envelope occurs only when the parent star is on the AGB, i.e., when the radius of the envelope is very large. In these two cases, HIP~114933~b and beta~Pic~c, we obtain 
$\frac{E_{\rm orbit}}{E_{\rm env}} = 0.11$ and $0.27$, respectively. This amount of energy is likely to inflate the envelope, therefore reducing the effective temperature. A lower effective temperature facilitates dust formation that in turn makes the mass loss process more susceptible to the dynamo activity and to the excitation of waves by the planet when it is deep in the envelope (and before it is destroyed). 

Overall, we take out findings to support the notion that massive planets, more massive than about Jupiter mass, can enhance the mass loss rate from RGB and AGB stars and shape their wind to a low degree of aspherical mass loss. 
    
\textbf{ACKNOWLEDGEMENTS}

We thank an anonymous referee for helpful comments. 
This research was supported by a grant from the Israel Science Foundation (769/20).

\textbf{DATA AVAILABILITY}

The data underlying this article will be shared on reasonable request to the corresponding author. 


\pagebreak


\begin{thebibliography}{}

\bibitem[Aguilera-G{\'o}mez et al.(2016)]{AguileraGomezetal2016} Aguilera-G{\'o}mez, C., Chanam{\'e}, J., Pinsonneault, M.~H., \& Carlberg, J.~K.\ 2016, \apj, 829, 127 

\bibitem[Akras et al.(2016)]{Akrasetal2016} Akras, S., Clyne, N., Boumis, P., Monteiro, H., Goncalves, D.~R., Redman, M.~P., \& Williams, S.\ 2016, \mnras, 457, 3409

\bibitem[Akras et al.(2020)]{Akrasetal2020} Akras, S., Monteiro, H., Aleman, I., Farias M.~A.~F., May D., \& Pereira C.~B.,\ 2020, \mnras, 493, 2238

\bibitem[Ali et al.(2016)]{Alietal2016} Ali, A., Dopita, M.~A., Basurah, H.~M., Amer, M.~A., Alsulami, R., \& Alruhaili, A.\ 2016, \mnras, 462, 1393

\bibitem[Aller et al.(2020)]{Alleretal2020} Aller, A., Lillo-Box, J., Jones, D., Miranda, L.~F., \& Barcel{\'o} Forteza, S., \ 2020, \aap, 635, A128. doi:10.1051/0004-6361/201937118

\bibitem[Barker et al.(2018)]{Barker2018} Barker, H., Zijlstra, A., De Marco, O., et al.\ 2018, \mnras, 475, 4504

\bibitem[Bear et al.(2021)]{Bearetal2021} Bear, E. Merlov, A., Arad, Y. \& Soker, 2021, arXiv:2106.00582

\bibitem[Bear et al.(2011)]{Bearetal2011} Bear, E., Soker, N., \& Harpaz, A.\ 2011, \apjl, 733, L44. doi:10.1088/2041-8205/733/2/L44

\bibitem[Berm{\'u}dez-Bustamante et al.(2020)]{BermudezBustamanteetal2020} Berm{\'u}dez-Bustamante, L.~C., Garc{\'\i}a-Segura, G., Steffen, W., et al.\ 2020, \mnras, 493, 2606

\bibitem[Bond, \& Ciardullo(2018)]{BondCiardullo2018} Bond, H.~E., \& Ciardullo, R.\ 2018, Research Notes of the American Astronomical Society, 2, 143

\bibitem[Bond et al.(2016)]{Bondetal2016} Bond, H.~E., Ciardullo, R., Esplin, T.~L., Hawley, S.~A., Liebert, J., \& Munari, U.\ 2016, \apj, 826, 139

\bibitem[Boyle(2018)]{Boyle2018PhDT} Boyle, L.~A.\ 2018, Ph.D. Thesis

\bibitem[Brown et al.(2019)]{Brownetal2019} Brown, A.~J., Jones, D., Boffin, H.~M.~J., et al.\ 2019, \mnras, 482, 4951. doi:10.1093/mnras/sty2986

\bibitem[Bujarrabal et al.(2018)]{Bujarrabaletal2018} Bujarrabal V., Castro-Carrizo A., Van Winckel H., Alcolea J., S{\'a}nchez Contreras C., Santander-Garc{\'\i}a M., Hillen M., 2018, A\&A, 614, A58

\bibitem[Carlberg et al.(2009)]{Carlbergetal2009} Carlberg, J.~K., Majewski, S.~R., \& Arras, P.\ 2009, \apj, 700, 832

\bibitem[Chamandy et al.(2021)]{Chamandyetal2021} Chamandy, L., Blackman, E.~G., Nordhaus, J., \& Wilson, E.\ 2021, \mnras, 502, L110. doi:10.1093/mnrasl/slab017

\bibitem[Chamandy et al.(2018)]{Chamandyetal2018a} Chamandy, L., Frank, A., Blackman, E.~G., et al.\ 2018, \mnras, 480, 1898

\bibitem[Chen et al.(2018)]{Chenetal2018} Chen Z., Blackman E.~G., Nordhaus J., Frank A., Carroll-Nellenback J., 2018, MNRAS, 473, 747

\bibitem[Chiotellis et al.(2016)]{Chiotellisetal2016} Chiotellis, A., Boumis, P., Nanouris, N., Meaburn, J., \& Dimitriadis, G.\ 2016, \mnras, 457, 9

\bibitem[Danehkar et al.(2018)]{Danehkaretal2018} Danehkar A., Karovska M., Maksym W.~P., Montez R., 2018, ApJ, 852, 87

\bibitem[Decin et al.(2020)]{Decinetal2020} Decin, L., Montarg{\`e}s, M., Richards, A.~M.~S., et al.\ 2020, Science, 369, 1497. doi:10.1126/science.abb1229

\bibitem[De Marco \& Soker(2011)]{DeMarcoSoker2011} De Marco, O., \& Soker, N.\ 2011, \pasp, 123, 402

\bibitem[Desmurs et al.(2019)]{Desmursetal2019} Desmurs, J.-F., Alcolea, J., Lindqvist, M., Bujarrabal, V., Soria-Ruiz, R., de Vicente, P., 2019,  arXiv:1905.07219

\bibitem[Frank et al.(2018)]{Franketal2018} Frank A., Chen Z., Reichardt T., De Marco O., Blackman E., Nordhaus J., 2018, Galax, 6, 113

\bibitem[Garc{\'{\i}}a-Rojas et al.(2016)]{GarciaRojasetal2016} Garc{\'{\i}}a-Rojas, J., Corradi, R.~L.~M., Monteiro, H., Jones, D., Rodriguez-Gil, P., \& Cabrera-Lavers, A.\ 2016, \apjl, 824, L27

\bibitem[Garc{\'\i}a-Segura et al.(2018)]{GarciaSeguraetal2018} Garc{\'\i}a-Segura, G., Ricker, P.~M., \& Taam, R.~E.\ 2018, \apj, 860, 19

\bibitem[Garc{\'\i}a-Segura et al.(2020)]{GarciaSeguraetal2020} Garc{\'\i}a-Segura, G., Taam, R.~E., \& Ricker, P.~M.\ 2020, \apj, 893, 150. doi:10.3847/1538-4357/ab8006

\bibitem[Garc{\'{\i}}a-Segura et al.(2014)]{GarciaSeguraetal2014}
Garc{\'{\i}}a-Segura, G., Villaver, E., Langer, N., Yoon, S.-C., \& Manchado, A.\ 2014, \apj, 783, 74

\bibitem[Glanz, \& Perets(2018)]{GlanzPerets2018} Glanz, H., \& Perets, H.~B.\ 2018, \mnras, 478, L12

\bibitem[Guo et al.(2017)]{Guoetal2017} Guo, J., Lin, L., Bai, C., \& Liu, J.,\ 2017, \apss, 362, 15. doi:10.1007/s10509-016-2978-7

\bibitem[Hegazi et al.(2020)]{Hegazietal2020} Hegazi, A., Bear, E., \& Soker, N.\ 2020, \mnras, 496, 612. doi:10.1093/mnras/staa1551

\bibitem[Hillwig(2018)]{Hillwig2018} Hillwig, T.\ 2018, Galaxies, 6, 85

\bibitem[Hillwig et al.(2016)]{Hillwigetal2016a} Hillwig, T.~C., Bond, H.~E., Frew, D.~J., Schaub, S.~C., \& Bodman, E.~H.~L.\ 2016, \aj, 152, 34

\bibitem[Iaconi et al.(2019)]{Iaconietal2019} Iaconi, R., Maeda, K., De Marco, O., Nozawa, T., \& Reichardt, T., 2019, \mnras, 489, 3334

\bibitem[Iaconi et al.(2020)]{Iaconi2020} Iaconi, R., Maeda, K., Nozawa, T., et al.\ 2020, \mnras, 497, 3166. doi:10.1093/mnras/staa2169

\bibitem[Iaconi et al.(2017b)]{Iaconietal2017} Iaconi, R., Reichardt, T., Staff, J., De Marco, O., Passy, J.-C., Price, D., Wurster, J., \& Herwig, F.\ 2017b, \mnras, 464, 4028

\bibitem[Jacoby et al.(2020)]{Jacobyetal2020} Jacoby, G.~H., Hillwig, T.~C., \& Jones, D.\ 2020, \mnras, 498, L114. doi:10.1093/mnrasl/slaa138

\bibitem[Jiang et al.(2020)]{Jiangetal2020} Jiang, C., Bedding, T.~R., Stassun, K.~G., et al.\ 2020, \apj, 896, 65. doi:10.3847/1538-4357/ab8f29

\bibitem[Jimenez et al.(2020)]{Jimenezetal2020} Jimenez, R., Gr{\r{a}}e J{\O}rgensen, U., \& Verde, L.\ 2020, \jcap, 2020, 027. doi:10.1088/1475-7516/2020/10/027

\bibitem[Jones(2020a)]{Jones2020CEE} Jones, D.\ 2020a, Reviews in Frontiers of Modern Astrophysics; From Space Debris to Cosmology, 123. doi:10.1007/978-3-030-38509-5\_5

\bibitem[Jones(2020b)]{Jones2020Galax} Jones, D.\ 2020b, Galaxies, 8, 28. doi:10.3390/galaxies8020028

\bibitem[Jones \& Boffin(2017b)]{JonesBoffin2017b} Jones, D., \& Boffin, H.~M.~J.\ 2017b, Nature Astronomy 1, 0117 

\bibitem[Jones et al.(2020a)]{Jonesetal2020RGB} Jones, D., Boffin, H.~M.~J., Hibbert, J., et al.\ 2020b, \aap, 642, A108. doi:10.1051/0004-6361/202038778

\bibitem[Jones et al.(2019)]{Jonesetal2019triple} Jones, D., Pejcha, O., \& Corradi, R.~L.~M.\ 2019, \mnras, 489, 2195. doi:10.1093/mnras/stz2293

\bibitem[Jones et al.(2016)]{Jonesetal2016} Jones, D., Wesson, R., Garc{\'{\i}}a-Rojas, J., Corradi, R.~L.~M., \& Boffin, H.~M.~J.\ 2016, \mnras, 455, 3263

\bibitem[Jones et al.(2021)]{Jonesetal2021} Jones, M.~I., Wittenmyer, R., Aguilera-G{\'o}mez, C., et al.\ 2021, \aap, 646, A131. doi:10.1051/0004-6361/202038555

\bibitem[Kervella et al.(2016)]{Kervellaetal2016} Kervella, P., Homan, W., Richards, A.~M.~S., Decin, L., McDonald, I., Montarg{\`e}s M., \& Ohnaka, K., \ 2016, \aap, 596, A92

\bibitem[Khouri et al.(2020)]{Khourietal2020} Khouri, T., Vlemmings, W.~H.~T., Paladini, C., et al.\ 2020, \aap, 635, A200. doi:10.1051/0004-6361/201834618

\bibitem[Kim et al.(2019)]{Kimetal2019} Kim, H., Liu, S.-Y., \& Taam, R.~E.\ 2019, \apjs, 243, 35

\bibitem[K{\H{o}}v{\'a}ri et al.(2019)]{Kovarietal2019} K{\H{o}}v{\'a}ri, Z., Strassmeier, K.~G., Ol{\'a}h, K., et al.\ 2019, \aap, 624, A83

\bibitem[Kramer et al.(2020)]{Krameretal2020} Kramer, M., Schneider, F.~R.~N., Ohlmann, S.~T., Geier, S., Schaffenroth, V., Pakmor, R., \& R{\"o}pke, F.~K., \ 2020, \aap, 642, A97. doi:10.1051/0004-6361/202038702

\bibitem[Lagos et al.(2021)]{Lagosetal2021} Lagos, F., Schreiber, M.~R., Zorotovic, M., G{\"a}nsicke, B.~T., Ronco, M.~P., \& Hamers, A.~S.,\ 2021, \mnras, 501, 676. doi:10.1093/mnras/staa3703

\bibitem[Leal-Ferreira et al.(2013)]{LealFerreiraetal2013} Leal-Ferreira, M.~L., Vlemmings, W.~H.~T., Kemball, A., \& Amiri, N.,\ 2013, \aap, 554, A134

\bibitem[Livio \& Soker(1988)]{LivioSoker1988} Livio, M., \& Soker, N.\ 1988, \apj, 329, 764

\bibitem[Lohev et al.(2019)]{Lohevetal2019} Lohev, N., Sabach, E., Gilkis, A., \& Soker, N.,\ 2019, \mnras, 490, 9. doi:10.1093/mnras/stz2593

\bibitem[L{\'o}pez-C{\'a}mara et al.(2019)]{LopezCamaraetal2019} L{\'o}pez-C{\'a}mara, D., De Colle, F., \& Moreno M{\'e}ndez, E.\ 2019, \mnras, 482, 3646

\bibitem[Lopez-Camara et al.(2020)]{LopezCamaraetal2020} Lopez-Camara, D., Moreno Mendez, E., \& De Colle, F.\ 2020, \mnras, 497, 2057. doi:10.1093/mnras/staa1983

\bibitem[MacLeod et al.(2018a)]{MacLeodetal2018a} MacLeod, M., Ostriker, E.~C., \& Stone, J.~M.\ 2018a, \apj, 863, 5. doi:10.3847/1538-4357/aacf08

\bibitem[MacLeod et al.(2018b)]{MacLeodetal2018b} MacLeod, M., Ostriker, E.~C., \& Stone, J.~M.\ 2018b, \apj, 868, 136
doi:10.3847/1538-4357/aae9eb

\bibitem[Madappatt et al.(2016)]{Madappattetal2016} Madappatt, N., De Marco, O., \& Villaver, E.\ 2016, \mnras,

\bibitem[Massarotti(2008)]{Massarotti2008} Massarotti, A.\ 2008, \aj, 135, 2287  

\bibitem[Miszalski et al.(2018)]{Miszalskietal2018PASA} Miszalski B., Manick, R., Miko{\l}ajewska, J., Van Winckel, H., I{\l}kiewicz K., 2018, PASA, 35, e027

\bibitem[Miszalski et al.(2019)]{Miszalskietal2019MNRAS487} Miszalski B., Manick R., Van Winckel H., Miko{\l}ajewska J., 2019, \mnras, 487, 1040

\bibitem[Munday et al.(2020)]{Mundayetal2020} Munday, J., Jones, D., Garc{\'\i}a-Rojas, J., et al.\ 2020, \mnras, 498, 6005. doi:10.1093/mnras/staa2753
   
\bibitem[Mustill \& Villaver(2012)]{MustillVillaver2012} Mustill, A.~J., \& Villaver, E.\ 2012, \apj, 761, 121

\bibitem[Nandez et al.(2014)]{Nandezetal2014} Nandez, J.~L.~A., Ivanova, N., \& Lombardi, J.~C., Jr.\ 2014, \apj, 786, 39

\bibitem[Nordhaus \& Blackman(2006)]{NordhausBlackman2006} Nordhaus, J., \& Blackman, E.~G.\ 2006, \mnras, 370, 2004

\bibitem[Nordhaus \& Spiegel(2013)]{NordhausSpiegel2013} Nordhaus, J., \& Spiegel, D.~S.\ 2013, \mnras, 432, 500

\bibitem[Nordhaus et al.(2010)]{Nordhausetal2010} Nordhaus, J., Spiegel, D.~S., Ibgui, L., Goodman, J., \& Burrows, A.\ 2010, \mnras, 408, 631

\bibitem[Ohlmann et al.(2016)]{Ohlmannetal2016} Ohlmann, S.~T., R{\"o}pke, F.~K., Pakmor, R., \& Springel, V.\ 2016, \apjl, 816, L9

\bibitem[Orosz et al.(2019)]{Oroszetal2019} Orosz, G., G{\'o}mez, J.~F., Imai, H., et al.\ 2019, \mnras, 482, L40

\bibitem[Passy et al.(2012)]{Passyetal2012} Passy, J.-C., De Marco, O., Fryer, C.~L., et al.\ 2012, \apj, 744, 52

\bibitem[Paxton et al.(2011)]{Paxtonetal2011} Paxton, B., Bildsten, L., Dotter, A., et al.\ 2011, \apjs, 192, 3

\bibitem[Paxton et al.(2013)] {Paxtonetal2013} Paxton, B., Cantiello, M., Arras, P., et al. 2013, \apjs, 208, 4

\bibitem[Paxton et al.(2015)]{Paxtonetal2015} Paxton, B., Marchant, P., Schwab, J., et al.\ 2015, \apjs, 220, 15

\bibitem[Paxton et al.(2018)]{Paxtonetal2018} Paxton, B., Schwab, J., Bauer, E.~B., et al.\ 2018, \apjs, 234, 34

\bibitem[Paxton et al.(2019)]{Paxtonetal2019} Paxton, B., Smolec, R., Schwab, J., et al.\ 2019, \apjs, 243, 10, doi:10.3847/1538-4365/ab2241

\bibitem[Privitera et al.(2016b)]{Priviteraetal2016II} Privitera, G., Meynet, G., Eggenberger, P., Vidotto, A.~A., Villaver, E., \& Bianda M.,\ 2016b, \aap, 593, A128. doi:10.1051/0004-6361/201628758

\bibitem[Privitera et al.(2016a)]{Priviteraetal2016I} Privitera, G., Meynet, G., Eggenberger, P., Vidotto, A.~A., Villaver, E., \& Bianda, M.,\ 2016a, \aap, 591, A45. doi:10.1051/0004-6361/201528044
  
\bibitem[Rao et al.(2018)]{Raoetal2018} Rao, S., Meynet, G., Eggenberger, P., Haemmerl{\'e}, L., Privitera, G., Georgy, C., Ekstr{\"o}m, S., et al.,\ 2018, \aap, 618, A18. doi:10.1051/0004-6361/201833107

\bibitem[Rasio \& Livio(1996)]{RasioLivio1996} Rasio, F.~A., \& Livio, M.\ 1996, \apj, 471, 366

\bibitem[Reichardt et al.(2019)]{Reichardtetal2019} Reichardt T.~A., De Marco O., Iaconi R., Tout C.~A., Price D.~J., 2019, MNRAS, 484, 631
 
\bibitem[Ricker \& Taam(2008)]{RickerTaam2008} Ricker, P.~M., \& Taam, R.~E.\ 2008, \apjl, 672, L41

\bibitem[Sabach \& Soker(2018a)]{SabachSoker2018a} Sabach, E., \& Soker, N.\ 2018a, \mnras, 473, 286

\bibitem[Sabach \& Soker(2018b)]{SabachSoker2018b} Sabach, E., \& Soker, N.\ 2018b, \mnras, 479, 2249

\bibitem[Sahai(2018)]{Sahai2018ASPC} Sahai, R.\ 2018, Science with a Next Generation Very Large Array, 403

\bibitem[Salas et al.(2019)]{Salasetal2019} Salas, J.~M., Naoz, S., Morris, M.~R., \& Stephan, A.~P.\ 2019, \mnras, 487, 3029. doi:10.1093/mnras/stz1515

\bibitem[Sandquist et al.(1998)]{SandquistTaam1998} Sandquist, E.~L., Taam, R.~E., Chen, X., Bodenheimer, P., \& Burkert, A.\ 1998, \apj, 500, 909

\bibitem[Schaffenroth et al.(2019)]{Schaffenrothetal2019} Schaffenroth, V., Barlow, B.~N., Geier, S., et al.\ 2019, \aap, 630, A80

\bibitem[Schneider et al.(2011)]{Schneideretal2011} Schneider, J., Dedieu, C., Le Sidaner, P., Savalle, R., \& Zolotukhin, I., \ 2011, \aap, 532, A79

\bibitem[Schreier et al.(2019)]{Schreieretal2019} Schreier, R., Hillel, S., \& Soker, N.\ 2019, \mnras, 490, 4748

\bibitem[Shiber et al.(2019)]{Shiberetal2019} Shiber, S., Iaconi, R., De Marco, O., Soker, N., 2019, \mnras, 488, 5615

\bibitem[Siess \& Livio(1999a)]{SiessLivio1999AGB} Siess, L., \& Livio, M.\ 1999a, \mnras, 304, 925

\bibitem[Soker(1992a)]{Soker1992} Soker, N.\ 1992a, \apj, 386, 190

\bibitem[Soker(1992b)]{Soker1992b} Soker, N.\ 1992b, \apj, 389, 628

\bibitem[Soker(1993)]{Soker1993} Soker, N.\ 1993, \apj, 417, 347

\bibitem[Soker(1996)]{Soker1996} Soker, N.\ 1996, \apjl, 460, L53

\bibitem[Soker(1998b)]{Soker1998AGB} Soker, N.\ 1998b, \mnras, 299, 1242

\bibitem[Soker(2000)]{Soker2000} Soker, N.\ 2000, Asymmetrical Planetary Nebulae II: From Origins to Microstructures, ASP Conference Series, Vol. 199. Edited by J. H. Kastner, N. Soker, and S. Rappaport., 199, 71

\bibitem[Soker(2001a)]{Soker2001a} Soker, N.\ 2001a, Astrophysics and Space Science Library, 181

\bibitem[Soker(2020)]{SokerWorkPlans2020} Soker, N.\ 2020,  Galaxies, 8, 26. doi:10.3390/galaxies8010026

\bibitem[Staff et al.(2016)]{Staffetal2016} Staff, J.~E., De Marco, O., Wood, P., Galaviz, P., \& Passy, J.-C.\ 2016, \mnras, 458, 832

\bibitem[Veras(2016)]{Veras2016} Veras, D.\ 2016, Royal Society Open Science, 3, 150571 

\bibitem[Vlemmings(2018)]{Vlemmings2018} Vlemmings, W.~H.~T.\ 2018, Contributions of the Astronomical Observatory Skalnate Pleso, 48, 187

\bibitem[Wesson et al.(2018)]{Wessonetal2018}  Wesson R., Jones D., Garc{\'\i}a-Rojas J., Boffin H.~M.~J., Corradi R.~L.~M., 2018, \mnras, 480, 4589

\bibitem[Wittenmyer et al.(2017)]{Wittenmyeretal2017} Wittenmyer, R.~A., Jones, M.~I., Zhao, J., et al.\ 2017, \aj, 153, 51. doi:10.3847/1538-3881/153/2/51

\bibitem[Zou et al.(2020)]{Zouetal2020} Zou, Y., Frank, A., Chen, Z., et al.\ 2020, \mnras, 497, 2855. doi:10.1093/mnras/staa2145

\end{thebibliography}
\end{document}